
\magnification=\magstep1
\font\titlefont=cmr10 scaled\magstep3  
\newbox\leftpage \newdimen\fullhsize \newdimen\hstitle
\newdimen\hsbody
\hoffset=0.0truein \voffset=0.20truein \hsbody=\hsize \hstitle=\hsize
\tolerance=1000\hfuzz=2pt \baselineskip=20pt plus 4pt minus 2pt
\global\newcount\meqno \global\meqno=1
\def\eqn#1#2{\xdef #1{(\the\meqno)}\global\advance\meqno by1
$$#2\eqno#1$$}
\global\newcount\refno \global\refno=1 \newwrite\rfile
\def\ref#1#2{{[\the\refno]}\nref#1{#2}}%
\def\nref#1#2{\xdef#1{[\the\refno]}%
\ifnum\refno=1\immediate\openout\rfile=refs.tmp\fi%
\immediate\write\rfile{\noexpand\item{[\the\refno]\ }#2}%
\global\advance\refno by1}

\def\figures{\centerline{{\bf Figure
Captions}}\medskip\parindent=40pt}
\def\fig#1#2{\medskip\item{Fig.~#1:  }#2}

    \def\frac#1#2{{#1\over#2}}
   
\nopagenumbers \hsize=\hsbody \pageno=0 ~ \vfill
\centerline{\titlefont Collapse of Randomly Self--Interacting
Polymers}
\bigskip
\centerline{{\sl Yacov Kantor}($^*$)($^{**}$)($^{***}$) and
{\sl Mehran Kardar}($^*$)}
\centerline{($^*$)Department of Physics, Massachusetts Institute of
Technology, Cambridge, MA 02139, U.S.A.}
\centerline{($^{**}$)Physics Department, Harvard University,
Cambridge MA 02138, U.S.A.}
\centerline{($^{***}$)School of Physics and Astronomy, Tel Aviv
University,
Tel Aviv 69 978, Israel\footnote{\dag}{Permanent address}}

\vfill\centerline{\bf ABSTRACT}\nobreak\medskip\nobreak\par{
We use complete enumeration and Monte Carlo techniques to
study self--avoiding walks with random nearest--neighbor interactions
described by $v_0q_iq_j$, where $q_i=\pm1$ is a quenched
sequence of ``charges'' on the chain.
For equal numbers of positive  and negative charges
($N_+=N_-$), the polymer with $v_0>0$ undergoes a
transition from self--avoiding behavior to a compact state
at a temperature $\theta\approx1.2v_0$. The collapse temperature
$\theta(x)$ decreases with the asymmetry $x=|N_+-N_-|/(N_++N_-)$
and vanishes at $x\approx0.6$.
For $v_0<0$, a $\theta$--point is present at all  $x$.
}
\vfill
\line{PACS. 36.20.--r Macromolecules and polymer molecules\hfill}
\line{PACS. 35.20.Bm General molecular conformation and symmetry;
stereochemistry\hfill}
\line{PACS. 64.60.--i General studies of phase transitions   \hfill}
\vfill\eject\footline={\hss\tenrm\folio\hss}

Polymers in a solvent are subject to both the hard core repulsions
between the monomers and somewhat longer range (e.g. van der Waals)
attractive interactions. At high temperatures $T$, the repulsive
interactions are dominant, and the  radius of gyration
(root--mean--squared size)  of the
polymer scales with the number  of monomers $N$ as $R_g\sim N^\nu$,
where $\nu\approx 0.588$ in three dimensions. On a discrete lattice,
the hard core interactions are usually modeled  by
self--avoiding walks (SAWs). The attractive interactions are then
included by introducing a negative energy for each pair of monomers
residing on neighboring lattice sites.
As temperature is lowered, a point is reached ($T=\theta$) where
the repulsive and attractive interactions effectively cancel and
the polymer behaves essentially as an ideal random walk\ref
\rdegideal{de Gennes P.G., J. Physique Lett. {\bf 36}, L--55
(1975).}\ref\rdup{Duplantier B.,  J. Physique Lett. {\bf 41},
L--409 (1980).}\ with $\nu={1/2}$. For $T<\theta$, the polymer
collapses into a  compact object  with $\nu={1/3}$. Numerous
Monte Carlo (MC) and exact  enumeration
studies of the $\theta$--point have been performed (see
Ref.\ref\rkrem{Kremer K., J. Phys. {\bf A15}, 2879 (1981).}\
and references therein).

The collapse transition for {\it heteropolymers} is particularly
interesting in view of its possible relation to the problem of
{\it protein  folding}\ref\rgen{See, e.g., Creighton T.E.,
{\it Proteins: Their Structure and Molecular Properties},
Freeman, San Francisco (1984).}\ref\rchan{Chan H.S. and
Dill K.A., Physics Today, p. 24, February  1993.}.
While models based on random heteropolymers significantly
oversimplify the specificity and complexity of real proteins,
they do bring in fresh  perspectives
from the statistical mechanics of random systems and
spin--glasses\ref\rstein{Stein D.L., Proc. Natl. Acad. Sci.
USA {\bf 82}, 3670 (1985);
Bryngelson J.D. and Wolynes P.G., Proc. Natl. Acad. Sci. USA
{\bf 84}, 7524 (1987).}--\nref
\rGOa{Garel T. and Orland H., Europhys. Lett. {\bf 6}, 307
(1988).}\nref
\rsha{Shakhnovich E.I. and Gutin A.M., Europhys. Lett.
{\bf 8}, 327 (1989).}
\nref\rkar{Karplus M. and Shakhnovich E., in {\it Protein Folding},
ed. by Creighton T.E., ch.4, p. 127, Freeman \& Co.,
New York (1992).}\ref\riori{Iori G., Marinari E.,
Parisi G., and Struglia M.V., Physica {\bf A185}, 98 (1992).}.
We shall consider polymers formed from two types of monomers,
labelled by $q_i=\pm 1$, and subject to a short range interaction
\eqn\eHI{ {\cal H}_I={1\over2}\sum_{i,j}q_iq_j{\cal V}\left({\bf
r}_i-{\bf r}_j\right),}
where ${\bf r}_i$ are the coordinates of the monomers.
Do these interactions modify the collapse transition
of heteropolymers described above?
A perturbative answer is provided by noting that\ref
\rKKepl{Kantor Y. and Kardar M., Europhys. Lett. {\bf 14}, 421
(1991).}\
that the relevance of such interactions in $d$ space dimensions
is controlled by the scaling exponent $y_I=1-d\nu$.
The inhomogeneities are marginal in the compact state
($\nu=1/d$),  and may cause the polymer to choose a
particular configuration dependent on the specifics of
the sequence. The statistical properties of such  states
are the subject of considerable current interest\ref
\rshakhevol{Shakhnovich E.I. and  Gutin A.M., Nature
{\bf 346}, 773 (1990).}\ref\rshakhenum{Shakhnovich E.I.
and Gutin A.M., J. Chem. Phys. {\bf 93},  5967 (1990).}.
Here we address the simpler question of how ${\cal H}_I$
affects the non--compact states. For ideal (non--interacting)
chains with $\nu=1/2$, ${\cal  H}_I$  is relevant in $d<2$:
In $d=1$ the polymer swells if like charges repel, and
collapses to a point, if they attract\rKKepl\ (sea also
Refs.\ref\rfost{Foster D.P., Vanderzande C.,  and Yeomans J.,
J. Stat. Phys. {\bf 69}, 857 (1992).}\ref\rstep{Stepanow S.,
Schultz M., and Sommer J.-U., Europhys. Lett.  {\bf 19},
273 (1992).}). For all other non-compact states ($\nu>1/d$)
{\it weak} interactions  are never relevant. However, we
shall demonstrate that strong interactions described by
${\cal H}_I$ do lead to the collapse of a self--avoiding chain.

We investigate a discretized model in which the only
homogeneous  interaction is the repulsion caused by the
constraints of self--avoidance. In the absence  of
randomness the polymer explores all SAWs with equal probabilities.
A short range random interaction is incorporated by assigning an
energy $v_0q_iq_j$ to every pair of  monomers, $i$ and $j$,  on
{\it neighboring} lattice sites.
We primarily  focussed on $v_0>0$, for which like charges repel
and opposite ones attract. This choice corresponds to strongly
screened Coulomb interactions.  The asymmetry in the
amount of positive and negative charges is measured
by \hbox{$x\equiv|N_+-N_-|/N$}.  We shall show that
for moderate values of $x$ the polymer undergoes a
$\theta$--transition and explore its properties. {\it Negative} $v_0$
describes a situation in which the like monomers attract, and unlike
ones repel. A mixture of hydrophobic and hydrophilic monomers would
exhibit such a tendency. In this case the $\theta$--transition is
present for all $x$.

We used exact enumeration to study the properties of
chains of up to $L=12$ steps ($N=13$ monomers). We examined all
spatial conformations of SAWs and all possible quenched sequences
of charges. Taking advantage of rotation and reflection symmetries,
as well as the degeneracies related to inverting the order of
the sequence or the signs of all charges, we reduced the
number of independent configurations to
4,162,866 for 2,080 sequences, i.e. a total of roughly
$8\times 10^9$ possibilities.  Since the number of cases grows
by an order of magnitude when $L$ is increased by unity, we could not
go beyond $L\le 12$. By comparison, the maximal chain--lengths
considered in enumeration studies of {\it compact} walks in
$d=3$\rshakhenum\  and $d=2$\ref\rcama{Camancho C. J. and
Thirumalai, Phys. Rev. Lett. {\bf 71}, 2505  (1993).}\
are about {\it twice longer}.  In the two dimensional case
the number  of spatial conformations is smaller by 2 to 3 orders
of magnitude, while in the  former case only a random sample of
quenches was used. Our results were supplemented by MC
simulations for $L\le95$, using the pivoting method\ref
\rlal{Lal M., Molec. Phys. {\bf 17}, 57 (1969); Madras N. and Sokal
A.D., J. Stat. Phys. {\bf 50}, 109 (1988).}.
(The method provides excellent equilibration at high and intermediate
temperatures, but does not permit equilibration of almost compact
structures\ref\rKKfuture{Kantor Y. and Kardar M., to be published.}.)

Exact enumeration provides a quite detailed picture of the energy
landscape. Fig.(1) depicts contours of the density of states in
the variables  $R_g^2$ and $E$ for (a) a homogeneous chain with
$q_i=1$ for all $i$, and (b) a particular quench that is
approximately neutral. All energy levels are proportional to
$v_0$ and Fig.(1) corresponds to choosing $v_0=+1$.
The corresponding densities for $v_0<0$ are obtained by simply
reflecting the figures around $E=0$. From such histograms, the
behavior as a function of temperature is deduced as follows:
At high $T$ the  system explores the region of highest density
and moves to the lowest energy states as $T\to0$. Both chains
are SAWs at $T\to\infty$ with  $R_g^2\approx2.4$. As $T$ is
reduced the uniformly charged chain expands; its scaling at
$T=0$ is still that of a SAW, as the only effect of the
potential is to exclude configurations with chain segments
on neighboring sites. This  effectively increases the range
of the excluded volume interaction and swells $R_g$ by a finite
factor. The random chain in Fig.(1b) behaves quite
differently, collapsing to a dense low energy configuration as
$T\to0$.

If the collapse of random chains is through a $\theta$ transition
similar to homogeneous polymers, we expect that at the transition
$R_g^2\sim L$ as in ideal random walks. This expectation is
confirmed in Fig.(2) which depicts the  $T$--dependence of
(quench--averaged) ${R_g^2}/L$ for $T$s ranging from 3 to 95.
(The quench average is exact for enumeration data, and obtained
from 20 quenches in the MC simulations.)
The curves intersect in the vicinity of the same point.
Since for $T<\theta$ the polymers are compact
(${R_g^2}/L\sim L^{-1/3}$), while for $T>\theta$ they are
expanded  SAWs (${R_g^2}/L\sim L^{0.17}$), the graphs
representing larger $L$s show steeper crossover. Despite a
slight $L$--dependence of the intersection points we can locate
the $\theta$--temperature with reasonable accuracy at
$T\approx 1.2v_0$,  but because of small chain lengths the
crossover exponents cannot be determined reliably.
We also examined the heat capacity per monomer  which
exhibits a sharp peak whose height increases with $L$. The
position of the peak is close to $T=0.5v_0$ for short chains
and moves towards higher $T$s for longer chains.

What happens to the collapse transition if the charges
on the chain are not exactly balanced ($x>0$)? Due to the
excess of repulsive interactions, the chains initially swell
on lowering temperature. Nevertheless, at sufficiently low
temperatures, they may find some compact configurations of
low energy. The $\theta$ temperature is expected to decrease
due to charge  imbalance. Fig.(3) depicts partial plots of
${R_g^2}/L$ vs. $T$ at  $x=0,\ {1/2}$, and $5/6$. The curves
for $x={1/ 2}$ are qualitatively similar to  $x=0$, but
with a lower  $\theta$ temperature. For $x={5/6}$, however,
the  curves do not intersect and there is no collapse
transition; the chain is a SAW at both high and low
temperatures. An approximate phase diagram in the $(x,T)$
plane is constructed in  Fig.(4): the $\theta$--temperature
decreases monotonically with $x$, finally disappearing at
$x\approx0.6$.
In the $T\rightarrow0$ and $x\rightarrow1$  limit, the action of
minority charges is to form loops in the SAW of majority charges.
This is somewhat reminiscent of  a model of SAWs with periodically
distributed ``stickers'' which attach in pairs\ref
\rcates{Cates M.E. and Witten T.A., Macromol. {\bf 19}, 732 (1986);
Baljon A.R.C., Macromol. {\bf 26}, 4339 (1993).}.
Unlike our model, the minority stickers are not randomly situated and
do not stick to the majority monomers. With such stickers, the
collapse transition apparently persists all the way to $x=1$.

The phase diagram is different for $v_0<0$, i.e. when
like charges attract. The limit of $x=1$ describes
a homogeneous polymer with nearest neighbor attractions.
This is precisely the model used to investigate the collapse of
uniform polymers, and is known to undergo a $\theta$--transition at
$T\approx3.7|v_0|$\ \ref\rjan{Janssens M. and  Bellemans A.,
Macromol. {\bf 9}, 303 (1976).}. At the other extreme, when $x=0$,
the situation somewhat resembles the case of $v_0>0$.
Although the model is {\it not symmetric} under changing the of
sign of $v_0$ which corresponds to changing the sign of $E$ in
Fig. (1), the energy landscape of Fig.(1b) appears
approximately unchanged by such a reflection. Therefore, it
is not surprising that there is again a collapse for $x=0$ at
$\theta\approx 1.2|v_0|$. We verified that the model with
$v_0<0$ does indeed have a collapse temperature  for all
values of $x$ which interpolates between the above two limits.

Most models of protein folding assume that a  {\it homogeneous}
short  range attraction leads to formation of a compact  phase
whose details are then determined by the particulars of the
sequence. In this work we have shown that even random
interactions with no ``built--in'' overall attraction can
create a compact state. Our polymers are not long enough to
determine the critical behavior at the collapse transition.
(Even for uniform polymers, chains with  $L>200$ are needed
to  determine exponents with any degree of accuracy.)
However, scaling arguments suggest that randomness is
irrelevant for $\theta$--chains, and that the collapse
transition should be in the  same universality as that of
uniform polymers. Some aspects of the model deserve
further study. It is plausible that a random polymer
initially collapses to a dense  liquid
state, followed by a glass transition upon further cooling.
(Studies that only enumerate compact states\rshakhenum\
appear to find such a transition.) It would also be
interesting to investigate the behavior of polymers close
to the point at which the $\theta$--transition disappears.
Detailed investigation of this point, however, requires
use of much longer chains ($L\approx 1000$) and methods
which permit  reliable equilibration of dense structures
at low temperatures.

This work was supported by the US--Israel BSF grant No. 92--00026,
by the NSF through No. DMR--87--19217  (at MIT's CMSE), DMR 91--15491
(at Harvard), and the PYI program (MK). YK would like to thank
D. Ertas for numerous discussions of the subject.

\vfill\eject\immediate\closeout\rfile
\centerline{{\bf References}}\bigskip
{\catcode`\@=11\escapechar=`  \input refs.tmp\vfill\eject}
\figures
\fig{1}{Contour plots of the number of states as a function of $R_g^2$
(in lattice units) and energy (in units of $v_0$) for a 10--step
polymer, in
(a) uniformly charged case ($N_+=11$),
and (b) for one quench in an almost neutral case ($N_+=6$, $N_-=5$).
The data
bins are of size 1 on the energy scale, and size $1/4$ on the $R_g^2$
scale. The smooth contour lines were created by interpolation at
levels 0.5, 33, 129, 513, 2049, 8000, 16000, 32000, 64000. The full
circles indicate the location of average $R_g^2$ as the temperature
changes
between 0 and $\infty$.}
\fig{2}{${R_g^2}$ (in lattice constants) divided by the
length of
the walk $L$ as a function
of temperature, for $N_+=N_-$. Dashed lines represent
the results of exact enumeration for $L=3$, 5, 7, 9, and 11,
in order of increasing
slope. Full squares and circles, open squares and circles, and open
triangles
represent
MC results for $L=9$, 13, 23, 47 and 95, respectively.}
\fig{3}{$R_g^2$ (in lattice constants) divided by the
length of the walk $L$, for $L=95$ (solid
lines), $L=47$ (dot--dashed lines), $L=23$ (dotted lines), and $L=11$
(dashed lines). The groups of lines
represent (from right to left) the values of $x=0,\ {1/2}$,
and $5/6$.}
\fig{4}{Phase diagram of a random polymer with $v_0>0$ in the plane
of
temperature (in units of $v_0$), and asymmetry $x$. Vertical bars
indicate
estimated uncertainties in the extrapolated values of the
$\theta$--points.}
\vfill\eject
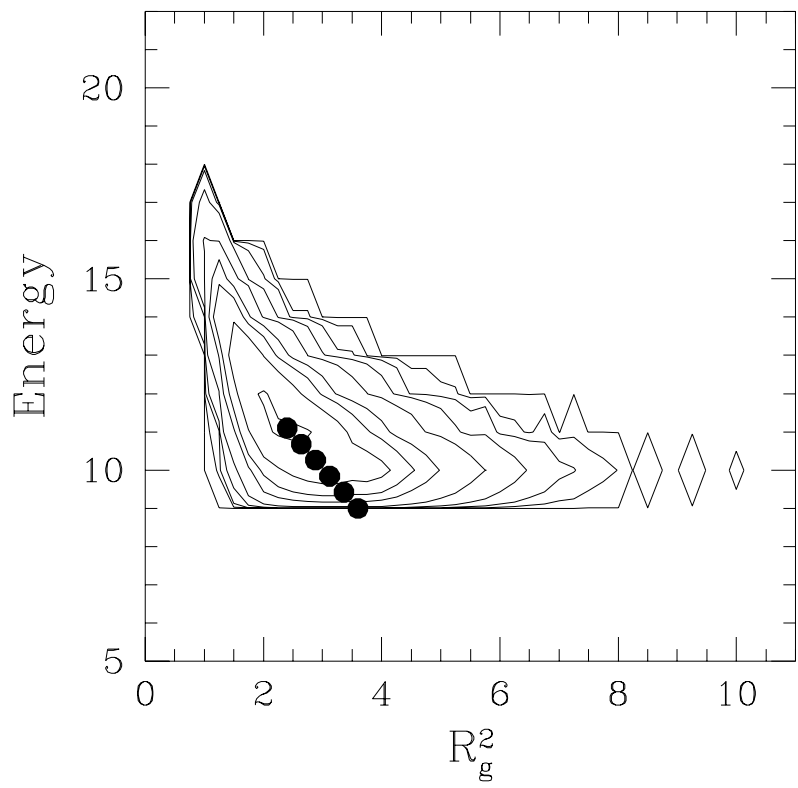
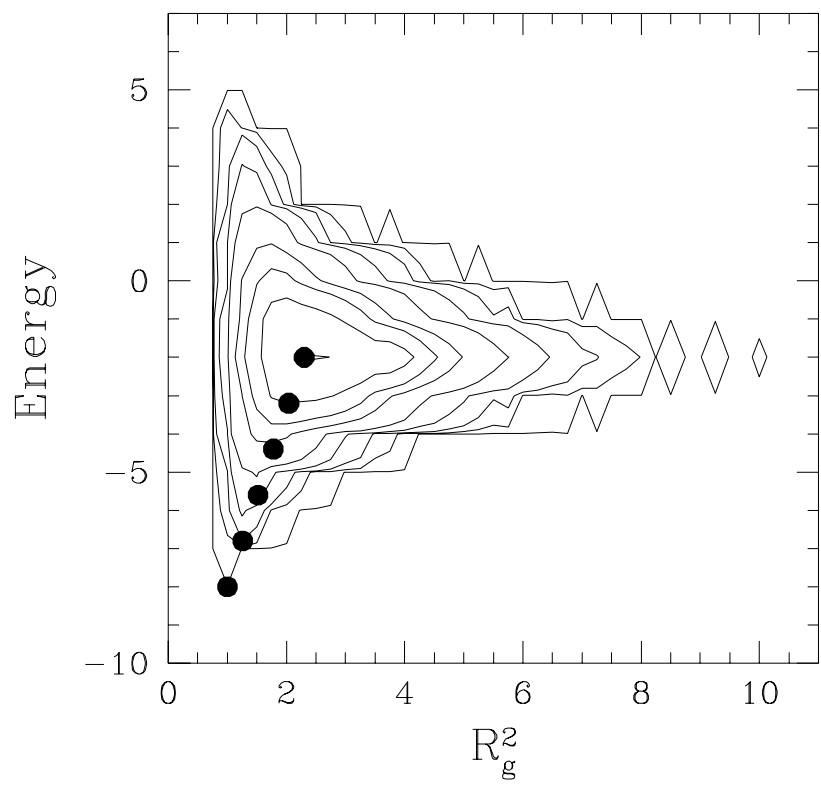
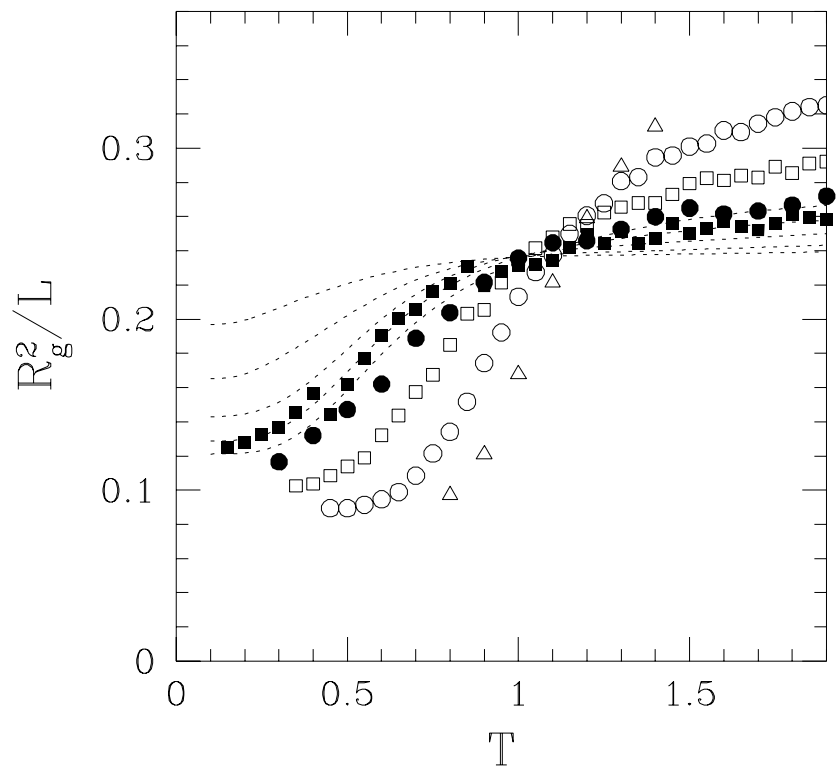
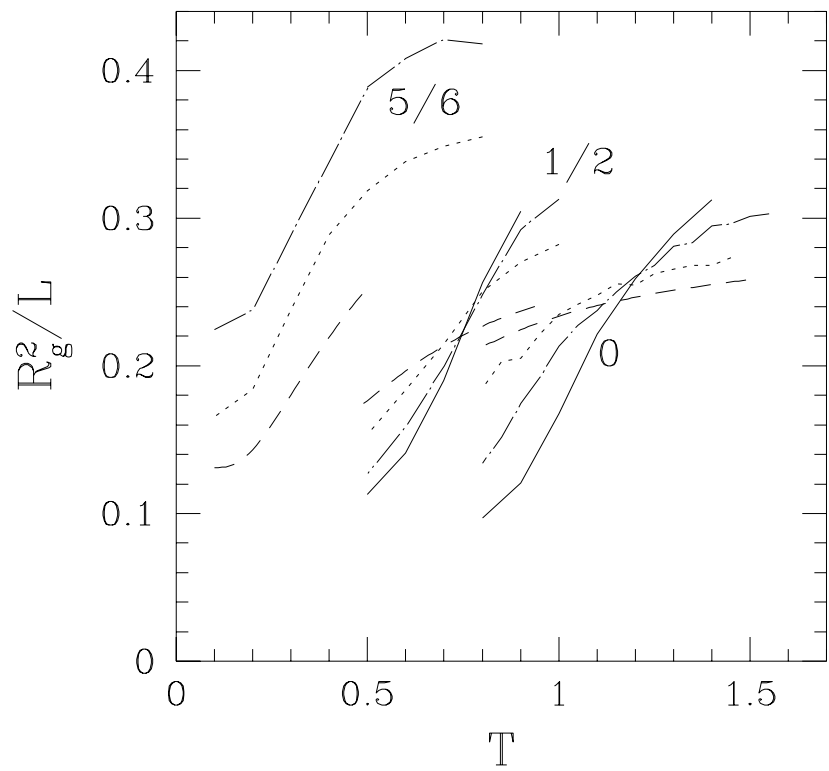
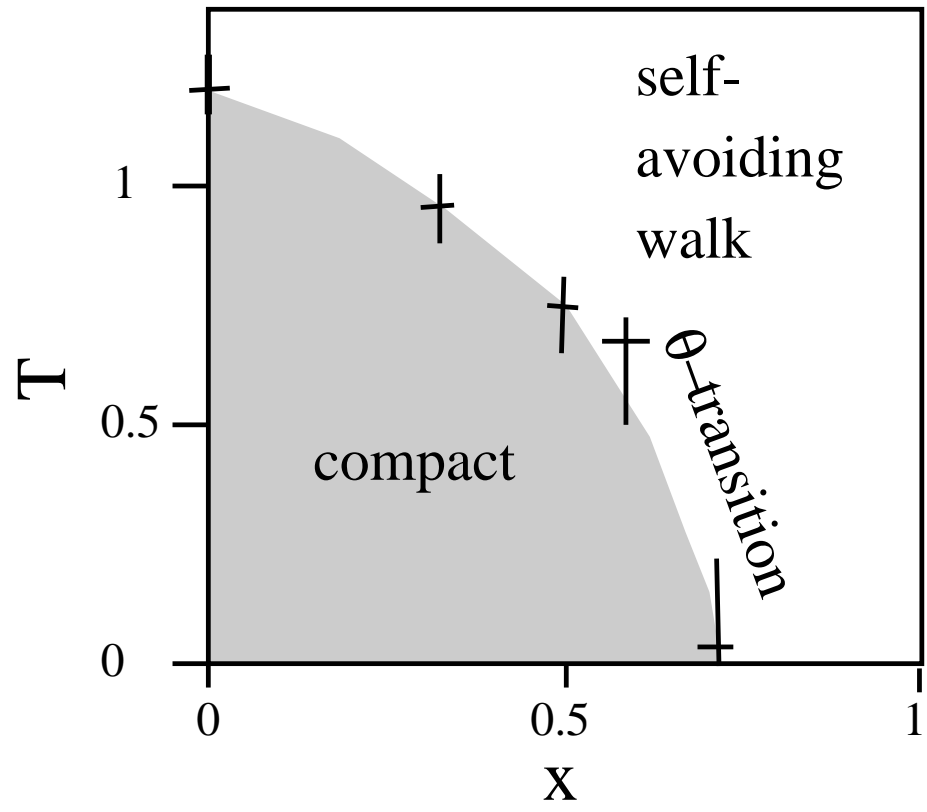
\end